\documentclass{webofc}

\usepackage[varg]{txfonts}   
\usepackage{hyperref}
\usepackage{url}
\usepackage{titlesec}

\hypersetup{colorlinks=true,citecolor=blue,urlcolor=blue,linkcolor=blue}
\begin{document}
\title{Probing bulk dynamics of the QGP with correlations and fluctuations}
%
%

\author{\firstname{Mesut} \lastname{Arslandok}\inst{1}\fnsep\thanks{\email{mesut.arslandok@cern.ch}}
}

\institute{Wright Lab, Physics Department, Yale University, New Haven, CT 06520, USA   }

\abstract{
 Ultrarelativistic heavy-ion collisions are considered ideal environments for exploring the QCD phase diagram and probing the properties of the QGP as functions of temperature and baryon chemical potential. At the highest energies, such as those reached at the Large Hadron Collider (LHC), and near vanishing baryon chemical potential, the transition from hadronic matter to the QGP is expected to occur as a smooth crossover. At larger baryon chemical potentials, this crossover may end at a critical point, beyond which the transition becomes first order. Locating this critical point remains a central goal of current and future beam energy scan programs at RHIC, the CERN SPS, and FAIR/GSI. Fluctuation and correlation measurements are widely used to probe the QCD phase structure, as they provide information on the system’s dynamical evolution and the nature of the phase transition at different regions of the phase diagram. This report presents an overview of recent experimental results across a broad energy range and discusses their implications for our current understanding of the QCD phase diagram.
}
\maketitle
%
\section{Introduction} \label{intro}
Quantum chromodynamics (QCD), the fundamental theory describing the strong interaction, predicts that when nuclear matter reaches sufficiently high energy densities, it undergoes a phase transition to a deconfined state of quarks and gluons, known as the quark–gluon plasma (QGP). To investigate the dynamics of this transition, and in particular the critical end point (CEP), an essential tool is the study of event-by-event fluctuations and correlations of global observables in the hadronic final state. At the CEP, a genuine thermodynamic singularity, susceptibilities diverge, and the order parameter exhibits long-wavelength fluctuations. Consequently, experimental signatures are expected to display non-monotonic variations as functions of collision energy or system size. The study of event-by-event fluctuations is therefore central to the energy scan programs at the CERN SPS, RHIC, and FAIR/GSI, which aim to map the QCD phase diagram and, in particular, to locate the CEP.

The fluctuations of conserved charges serve as sensitive probes of the QCD equation of state and can be directly connected to thermodynamic susceptibilities, which are calculable within the framework of lattice QCD (LQCD)~\cite{HotQCD}. Quark-number susceptibilities ($\chi$) are defined as derivatives of the reduced QCD pressure ($P/T^{4}$) with respect to the reduced chemical potentials ($\hat{\mu}=\mu/T$) associated with conserved charges, such as electric charge $Q$, baryon number $B$, and strangeness $S$.
Experimentally, these susceptibilities are studied through cumulants of net-charge distributions, $\kappa_{n}(N)$, with $\chi_{n} = (1/VT^{3})\kappa_{n}(N)$~\cite{bridging}. Alternatively, factorial cumulants are also used, as they are linear combinations of cumulants and, at a given order $n$, exhibit different sensitivities to phase transitions~\cite{cumulants}. In lattice QCD, the volume $V$ and temperature $T$ are fixed external parameters, whereas in experiments they evolve dynamically with time and are not directly measurable. Therefore, experimental studies are typically performed in terms of ratios of cumulants, which remove the explicit dependence on $V$ and $T$. 
These ratios are then compared to the corresponding cumulants of the Skellam distribution, defined as the probability distribution of the difference between two statistically independent Poisson variables. For instance, at LHC energies where $\mu_{\rm B}$ is close to 0, cumulant ratios for the Skellam distribution are equal to 0 for odd orders and 1 for even orders. Any deviation from this Skellam baseline indicates the presence of dynamical correlations or nontrivial physical effects. One should also note that conserved charges, such as the net baryon number, cannot be measured directly, since identifying all baryons is technically not feasible. Therefore, proxies such as the net-proton number are used instead. 
\begin{figure}[h]
\centering
\includegraphics[width=\linewidth]{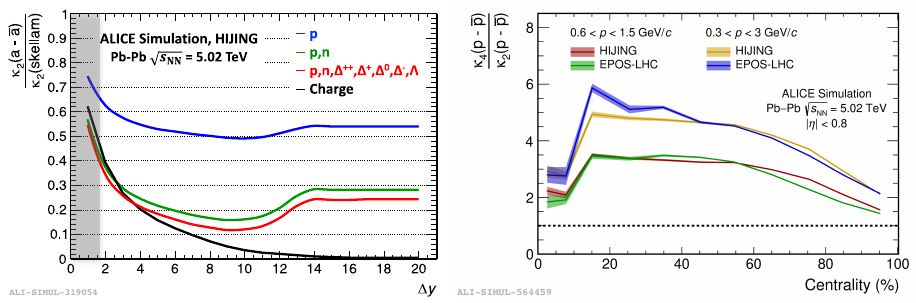}
\caption{(Color online) Left: HIJING~\cite{hijing} calculations of the rapidity dependence of second-order cumulants of net baryons (blue, green, red) and net electric charge (black), normalized to Skellam expectations. The grey band indicates the ALICE acceptance. Right: Centrality dependence of the $\kappa_{4}/\kappa_{2}$ ratio of net protons from HIJING~\cite{hijing} and EPOS-LHC~\cite{epos} for two momentum acceptances.}
\label{figure_1}       
\end{figure}

Figure~\ref{figure_1} (left) shows the rapidity dependence of the second-order cumulants of net baryons and net electric charge, normalized to the corresponding Skellam baseline. The deviation from unity is mainly caused by baryon number and electric charge conservation. As expected, in full phase space, the electric charge does not fluctuate and converges to zero, whereas the net-proton cumulants saturate around 0.5 near the beam rapidity, leading to an increase around rapidity 7. The growing number of contributing baryons at forward rapidities further enhances the deviation from the Skellam baseline. It should be noted that beam protons do not affect the measurements within the kinematic acceptance of ALICE; however, they play a crucial role in interpreting results at lower energies at RHIC, SPS, SIS18, and FAIR.

\section{Recent progress on experimental challenges} \label{sec-2}
Measurements of cumulants are experimentally very challenging because, unlike particle spectra measurements, not only the mean but also the higher-order moments of the multiplicity distributions must be corrected for experimental effects such as efficiency losses and event pile-up, both of which have been addressed recently~\cite{efficiency, pileup}. The two remaining major experimental challenges are volume fluctuations~\cite{bridging} and particle identification.

\subsection{Mixed event technique for volume fluctuations}
As mentioned earlier, cumulant ratios remove the explicit volume dependence but not the effects of volume fluctuations. These effects are illustrated in figure~\ref{figure_1} (right), which shows EPOS and HIJING model calculations for the fourth-to-second order cumulant ratio of the net-proton number in two different momentum ranges. The measured values lie systematically above the Skellam baseline (unity) and show a strong dependence on particle multiplicity within the kinematic acceptance, whereas the LQCD predicts values of about 0.75~\cite{latticekappa4}. Additionally, a clear dependence on the centrality bin width is observed when going from a 5\% to a 10\% range for centralities above 10\%. 

\begin{figure}
\centering
\includegraphics[width=\linewidth]{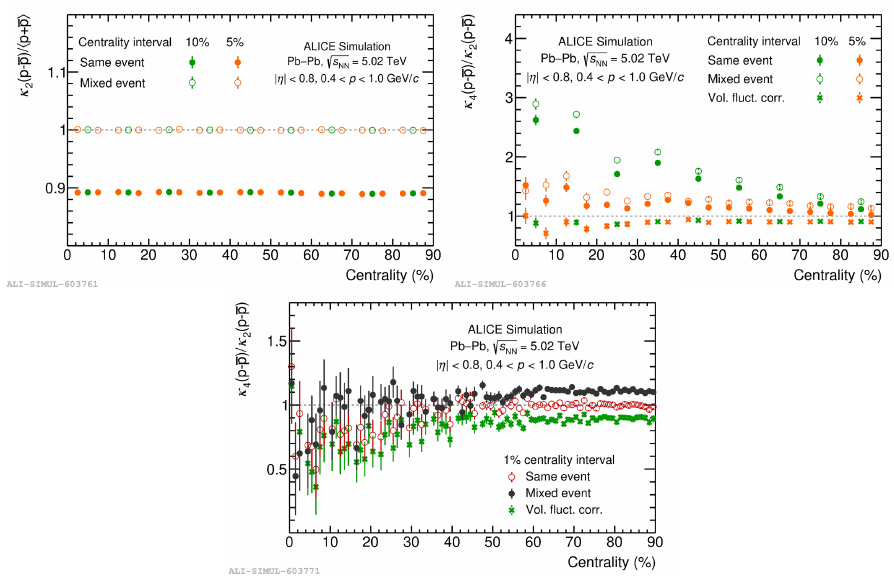}
\caption{(Color online) HIJING model calculations of the centrality dependence of net-proton cumulants for different bin widths, shown for same-event and mixed-event samples. The top left panel displays the normalized second-order cumulants relative to the Skellam expectation, while the top right and bottom panels show the ratio $\kappa_{4}/\kappa_{2}$ before and after volume-fluctuation corrections for 5–10\% and 1\% bins, respectively~\cite{ilya}.}
\label{figure_2}       
\end{figure}

The ALICE Collaboration studied the impact of centrality bin width using HIJING simulations for the second- and fourth-order cumulant ratios, as shown in figure~\ref{figure_2}~\cite{ilya}. The volume fluctuations were corrected using the mixed-event technique~\cite{mixedevent1, mixedevent}, where mixed events are classified with respect to event multiplicity and event-shape classes to account for potential biases from event geometry. At the second-order level (figure~\ref{figure_2}, top left), the mixed-event technique washes out all correlations and reproduces the Skellam baseline, confirming that it isolates the impact of volume fluctuations. Notably, the difference between corrected (green markers) and uncorrected (red markers) results with a 1\% centrality bin width indicates that volume fluctuations are not entirely eliminated by narrowing the bin width. This shows that the commonly used centrality bin width correction (CBWC) method~\cite{cbwc} can suppress, but not completely remove, the effect of volume fluctuations.

\subsection{Fuzzy Logic for particle identification}
\begin{figure}
\centering
\includegraphics[width=\linewidth]{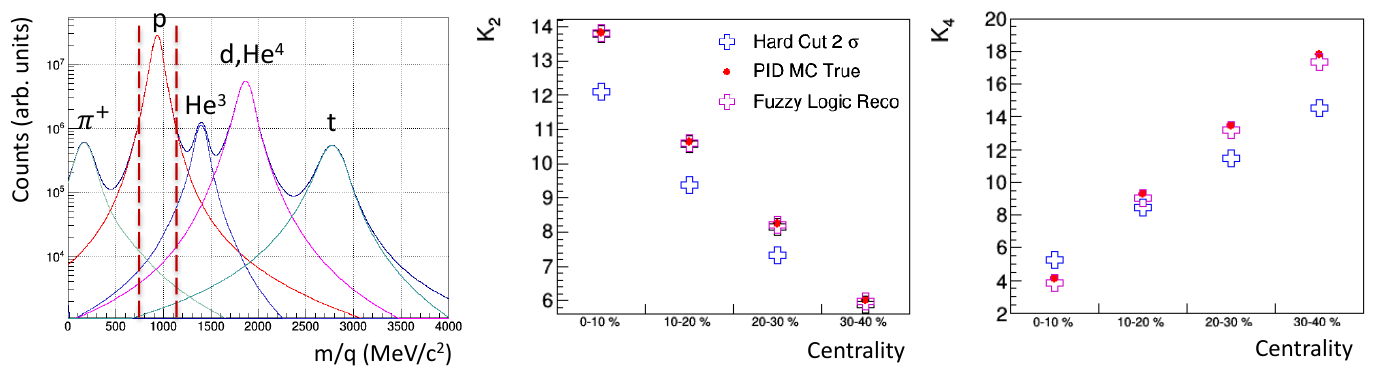}
\caption{(Color online) Left: Mass spectrum measured by the HADES collaboration with Gaussian fits to individual particle species; dashed lines indicate the $2\sigma$ selection window around the proton mean. Middle and right: second- and fourth-order proton cumulants from the particle-counting and Fuzzy Logic methods in full simulation of the SMASH transport model~\cite{smash}, compared to the true cumulants obtained from MC truth.}
\label{figure_3}       
\end{figure}

The standard approach to obtaining the moments, and thus the cumulants, of particle multiplicity distributions is to count the number of particles event by event with precise particle identification, since no reliable method exists to correct for sample contamination. However, this approach faces difficulties such as incomplete identification due to overlapping mass distributions as illustrated in figure~\ref{figure_3} (left)~\cite{Nabroth:2025qm}. These can be mitigated either by restricting the analysis to suitable phase-space regions (see red dashed lines in figure~\ref{figure_3}, left) or by incorporating additional detector information. Both strategies reduce the overall phase-space coverage and detection efficiency and still do not completely eliminate contamination.

A recently developed technique, known as Fuzzy Logic~\cite{fuzzy}, overcomes this limitation by adopting a probabilistic approach based on inclusive mass distributions (figure~\ref{figure_3}, left). It determines the moments of the multiplicity distributions through an unfolding procedure. The performance of Fuzzy Logic compared to traditional particle counting, as well as to the true cumulants obtained from Monte Carlo truth information, is shown for the second and fourth-order cumulants in figure~\ref{figure_3} (middle and right). As seen, even a small degree of contamination in the particle-counting method leads to large deviations from the true values, whereas Fuzzy Logic achieves excellent agreement.

\section{Status near the QCD crossover} \label{sec-3}
One of the central questions in high-energy physics is whether the QGP phase can be created in small systems such as pp collisions, where high-multiplicity events are already known to exhibit signs of collective behavior. However, it remains inconclusive whether this collective behavior originates from the QGP or from other dynamical effects. Cumulant ratios can be used to further test for possible critical fluctuations associated with a QGP phase transition. The LQCD predicts that a negative value of the sixth-order susceptibility (around -0.5) would indicate a crossover transition~\cite{latticekappa4}.
\begin{figure}[h]
\centering
\includegraphics[width=\linewidth]{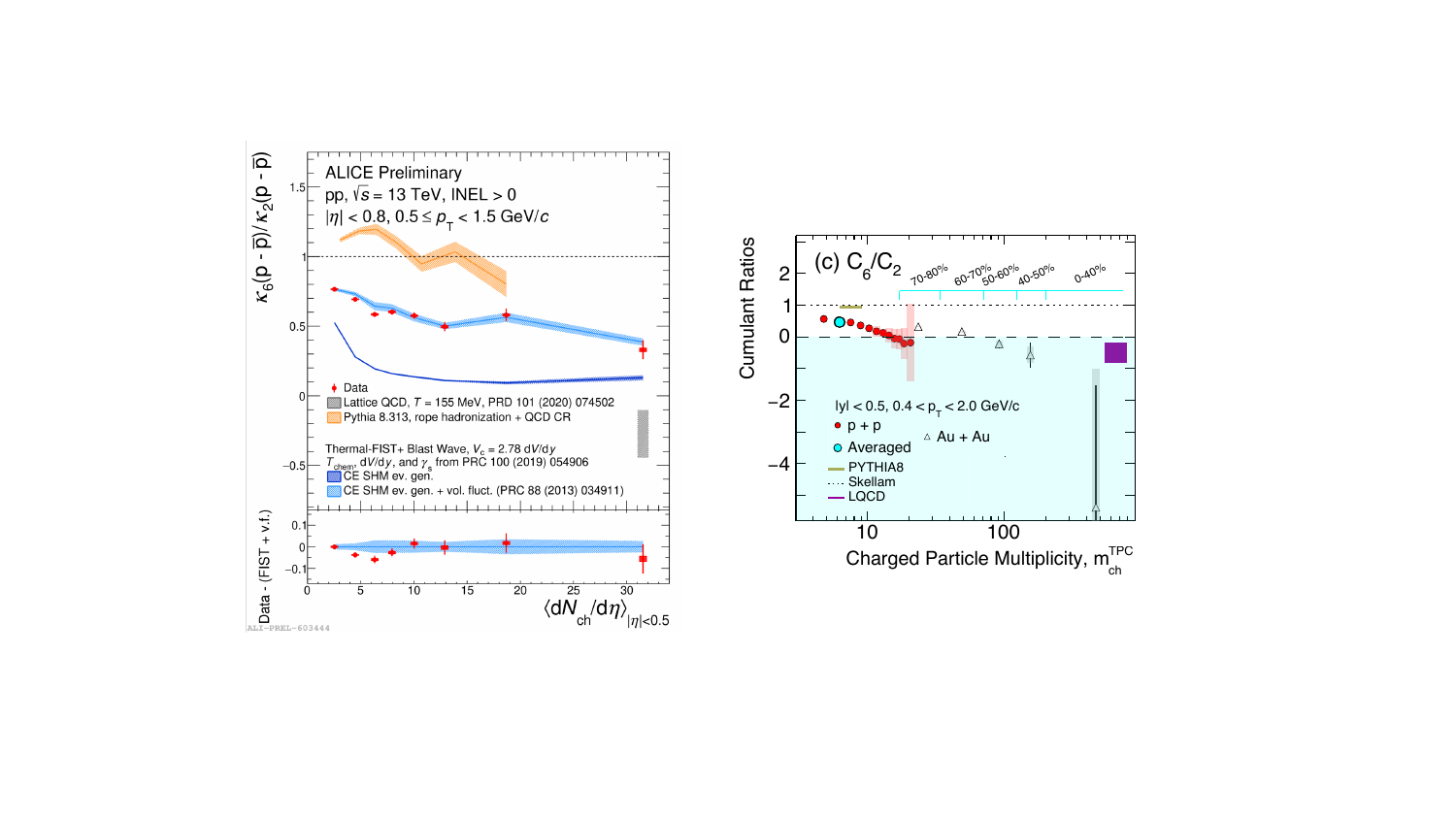}
\caption{Sixth- to second-order cumulant ratios of the net-proton number in pp collisions measured by ALICE~\cite{ilya} (left) and in pp and Pb--Pb collisions measured by STAR~\cite{starpp} (right). ALICE results are compared to PYTHIA and Thermal-FIST model calculations, while STAR results are compared to PYTHIA only.}
\label{figure_4}      
\end{figure}

While A--A collisions demand extensive statistics due to their high multiplicities, the comparatively low multiplicities in pp collisions make it easier to achieve the required experimental precision. The ALICE and STAR~\cite{starpp} collaborations have reported measurements of the sixth-to-second order cumulant ratios of net protons at 13 TeV and 200 GeV, respectively (see figure~\ref{figure_4}). The ALICE results were compared to Thermal-FIST~\cite{fist} calculations that account for baryon number conservation and volume fluctuation effects, showing consistency with the model as a function of multiplicity and thus no indication of critical behavior. STAR applied the CBWC method~\cite{cbwc} to correct for volume fluctuations and observed a clear decrease in the ratio with increasing multiplicity. However, due to the large experimental uncertainties and the absence of a subtraction for baryon number conservation effects, it remains inconclusive whether these results reflect signs of critical behavior.

Overall, neither dataset provides clear evidence for QGP-related signals. It should be noted, however, that the event multiplicities reached at the corresponding energies may still be insufficient. Future analyses of pp collisions at higher multiplicities, particularly for different event-shape classes such as isotropic collisions, and employing the event-mixing technique~\cite{mixedevent} to correct for volume fluctuations, could offer promising opportunities.
\begin{figure}[h]
\centering
\includegraphics[width=0.95\linewidth]{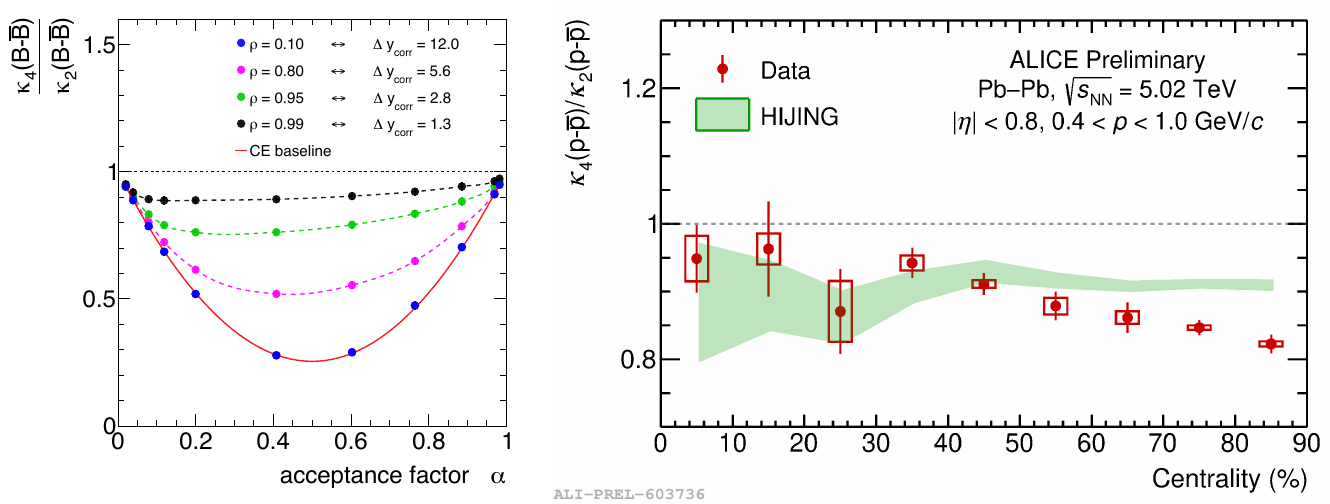}
\caption{Left: Effect of local (dashed lines) and global (solid red line) baryon number conservation on the ratio of fourth- to second-order cumulants of the net-baryon number as a function of rapidity~\cite{metropolis}. Right: Centrality dependence of fourth- to second-order cumulants of the net-proton number measured by the ALICE Collaboration in Pb--Pb collisions~\cite{ilya}. Results are compared to HIJING model calculations. Both data and model results are corrected for volume fluctuation effects using the mixed-event technique~\cite{mixedevent}.}
\label{figure_5}       
\end{figure}

The results for higher-order cumulants in A--A collisions beyond fourth order are statistically insufficient at both LHC and RHIC energies (see figure~\ref{figure_4}, right, triangle markers). For the fourth order, however, both the ALICE and STAR collaborations have reported promising results, showing deviations from the Skellam baseline and approaching the predictions of LQCD, although a direct comparison is not straightforward, as the net-proton number serves only as a proxy for the net-baryon number. Moreover, a fair comparison requires establishing the proper baseline, which must account for effects such as baryon number conservation, which strongly depends on the correlation length between baryon–antibaryon production. This effect is modeled in figure~\ref{figure_5} (left) within the Canonical Ensemble (CE) formulation, where an infinite correlation length corresponds to global conservation, while a finite correlation length represents local conservation effects~\cite{metropolis}. Larger deviations indicate longer correlation lengths and, consequently, earlier production. ALICE results, consistent within uncertainties with STAR data (see figure~\ref{figure_6}, right), for the fourth-to-second order net-proton cumulant ratio (figure~\ref{figure_5}, right) exhibit a clear decreasing trend toward more peripheral events. Understanding this trend in the context of a proper baseline that incorporates dynamical effects such as collectivity, resonance decays, and baryon number conservation is crucial for a quantitative comparison to LQCD expectations.

\section{Search for the critical end point} \label{sec-4}
\begin{figure}[h]
\centering
\includegraphics[width=\linewidth]{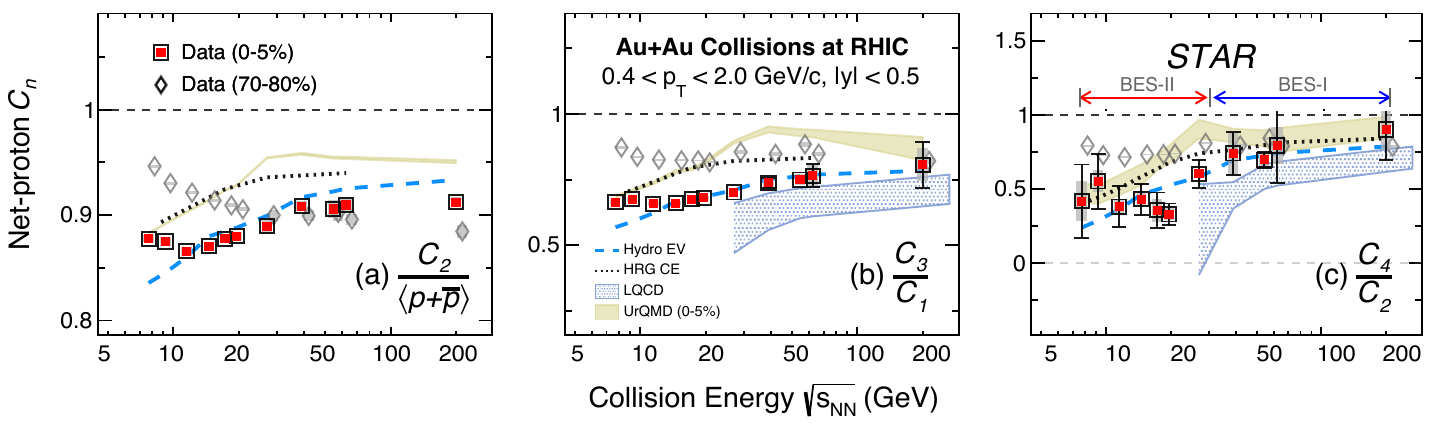}
\caption{Net-proton cumulant ratios measured by the STAR Collaboration in the Beam Energy Scan I and II programs at RHIC~\cite{starscan}. Left: second to first, middle: third to first, and right: fourth to second. Theoretical calculations are from a hydrodynamical model~\cite{hydroev} (blue dashed line), a thermal model with canonical treatment of baryon number (black dashed line), UrQMD transport model~\cite{urqmd} (brown band), and LQCD (light blue band).}
\label{figure_6}       
\end{figure}

\begin{figure}
\centering
\includegraphics[width=0.9\linewidth]{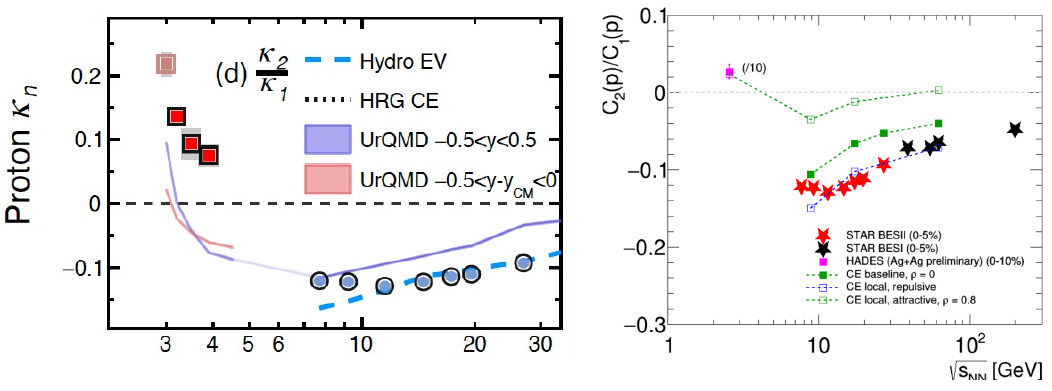}
\caption{Second-to-first-order proton factorial cumulant ratios measured by the STAR Collaboration in fixed-target mode (red boxes) and in the BES I and II programs (blue circles), compared to results from the HADES Collaboration.}
\label{figure_7}       
\end{figure}

The STAR Collaboration has reported cumulant and factorial cumulant results of the proton number from the fixed-target program as well as from BES-I and BES-II~\cite{starscan}. A clear deviation from the CE baseline with excluded volume, which accounts for the repulsive interaction between protons, is observed below 10 GeV (see Fig.~\ref{figure_6}). This deviation is consistent across the second to fourth-order cumulants. These results are complemented by fixed-target measurements from STAR~\cite{Zachary:2025qm} and the HADES~\cite{Nabroth:2025qm} experiment below 5 GeV, where a clear change to positive values is observed (see Fig.~\ref{figure_7}). This change of sign has recently been interpreted in ref.~\cite{anarbengt} as a transition from repulsive to attractive interactions, which may signal a first-order chiral phase transition. 

Although the UrQMD model also reproduces the change from negative to positive values, it is difficult to isolate the underlying physics mechanism responsible for this behavior within the model. Moreover, it should be noted that the STAR data employ the CBWC method for volume-fluctuation correction, which, as discussed above, may undercorrect. A reanalysis of the data using the mixed-event technique could therefore strengthen the physics interpretation of these results.

\section{Summary} \label{sec-5}
With the resolution of two major experimental challenges —volume fluctuations and particle misidentification —through the application of mixed-event and Fuzzy Logic methods, respectively, comparisons between experimental data and theoretical predictions have become more reliable than ever. Although current experimental results show no indication of QGP-like signals in pp collisions, future studies with higher event multiplicities, incorporating event-shape dependence and the state-of-the-art correction techniques, may still provide opportunities to search for QGP formation in small systems. The recently collected O–O and Ne–Ne data at the LHC are especially promising in this regard. 

For exploring the crossover regime, the fourth-order cumulant, measured with reasonable experimental precision by the ALICE and STAR collaborations, appears particularly promising. However, more differential studies relative to a properly defined canonical ensemble baseline are required. The RHIC energy-scan program, spanning fixed-target mode to BES-II, has revealed promising signals consistent with a possible first-order phase transition, manifested as a change from repulsive to attractive interactions at collision energies below 5 GeV. The results from HADES are in line with these findings. Confirmation of this observation with an independent observable, such as intermittency, which is characterized by increased fluctuations in local phase space in either momentum or coordinate space, would provide strong support for a potential discovery of the CEP, even though no convincing signals have yet been observed from intermittency in momentum space~\cite{na61inter,starinter}. As for the future, the upcoming ALICE 3 experiment at the LHC and the CBM experiment at GSI/FAIR offer excellent prospects for high-precision exploration of the QCD phase diagram, covering both the high- and low-$\mu_B$ regions.


\begin{thebibliography}{}
%
%
\bibitem{HotQCD}
HotQCD Collaboration, Phys. Rev. D 101, 074502 (2020).

\bibitem{bridging}
P. Braun-Munzinger, A. Rustamov, and J. Stachel, Nucl. Phys. A 364 960 114 (2017).

\bibitem{cumulants}
A. Bzdak, V. Koch, and N. Strodthoff, Phys. Rev. C 95 054906 (2017).

\bibitem{hijing}
M. Gyulassy and X.-N. Wang, Comput. Phys. Commun. 83 441 307 (1994).

\bibitem{epos}
T. Pierog, I. Karpenko, J. M. Katzy, E. Yatsenko, and K. Werner, Phys. Rev. C 92 460 034906 (2015).

\bibitem{efficiency}
T. Nonaka, M. Kitazawa, and S. Esumi, Nucl. Instrum. Meth. A 906 10-17 (2018).

\bibitem{pileup}
T. Nonaka, M. Kitazawa, S. Esumi, Nucl. Instrum. Meth. A 984 164632 (2020).

\bibitem{latticekappa4}
A. Bazavov et.al., Phys. Rev. D 95 054504 (2017).

\bibitem{ilya}
I. Fokin, (ALICE Collaboration), these proceedings.

\bibitem{mixedevent}
R. Holzmann, V. Koch, A. Rustamov, and J. Stroth, Nucl. Phys. A 1050 122924 (2024).

\bibitem{mixedevent1}
A.~Rustamov, J.~Stroth and R.~Holzmann, Nucl. Phys. A 1034 122641 (2023).

\bibitem{cbwc} 
X. Luo, J. Xu, B. Mohanty, and N. Xu, J. Phys. G 40, 105104 (2013).

\bibitem{smash}
SMASH Collaboration, Phys. Rev. C 94, 054905 (2016).

\bibitem{fuzzy}
A. Rustamov, Phys. Rev. C 110 6, 064910 (2024).

\bibitem{Nabroth:2025qm}
M. Nabroth, (HADES Collaboration), these proceedings.

\bibitem{Zachary:2025qm}
Z. Sweger, (STAR Collaboration), these proceedings.
  
\bibitem{smallsystem}
J. F. Grosse-Oetringhaus and U. A. Wiedemann, CERN-TH-2024-110 (2024).

\bibitem{starpp}
STAR Collaboration, Phys. Lett. B 857 138966 (2024).

\bibitem{fist}
V. Vovchenko and H. Stoecker, Comput. Phys. Commun. 244, 295 (2019).

\bibitem{metropolis}
P. B.-Munzinger, K. Redlich, A. Rustamov, and J. Stachel, JHEP 08 113 (2024).

\bibitem{starscan}
STAR Collaboration, Phys. Rev. Lett. 135 14, 142301 (2025).

\bibitem{hydroev}
V. Vovchenko, V. Koch, and C. Shen, Phys. Rev. C 105, 014904 (2022).

\bibitem{urqmd}
S. A. Bass et al., Prog. Part. Nucl. Phys. 41, 255 (1998).

\bibitem{anarbengt}
B. Friman, K. Redlich, and A. Rustamov, arXiv:2508.18879 (2025).

\bibitem{na61inter}
NA61/SHINE Collaboration, Andrzej Rybicki, EPJ Web Conf. 316 01008 (2025).

\bibitem{starinter}
STAR Collaboration, Phys. Lett. B 845 138165 (2023).



\end{thebibliography}
%
%

\end{document}